\documentclass[twocolumn,aps,superscriptaddress]{revtex4-1}
\usepackage{palatino}
\usepackage[latin1]{inputenc}
\usepackage{epsf}
\usepackage{amsmath,amssymb}
\usepackage{calc}

\usepackage{float}

\usepackage[usenames,dvipsnames]{xcolor}

\usepackage{graphicx}
\usepackage{hyperref}

\newcommand{\modified}{}

\newcommand{\ben}{\begin{equation}}
\newcommand{\een}{\end{equation}}
\newcommand{\bea}{\begin{eqnarray}}
\newcommand{\eea}{\end{eqnarray}}

\def\sss{\scriptscriptstyle\rm}

\def\subc{_{\sss C}}

\def\subxc{_{\sss XC}}

\def\1s{_{1,\sss S}}
\def\2s{_{2,\sss S}}
\def\x{_{\sss X}}
\def\c{_{\sss C}}
\def\s{_{\sss S}}
\def\xc{_{\sss XC}}

\def\Hxc{_{\sss HXC}}

\def\H{_{\sss H}}

\def\ext{_{\rm ext}}
\def\S{^{\rm S}}

\def\br{{\bf r}}
\def\bR{{\bf R}}

\def\bb{{\bf b}}

\begin{document}
 \title{ Non-Adiabatic Approximations in Time-Dependent Density Functional Theory: Progress and Prospects}
 \author{Lionel Lacombe}
 \affiliation{Laboratoire des Solides Irradi{\'e}s, {\'E}cole Polytechnique, Institut Polytechnique de Paris, F-91128 Palaiseau, France}
 \author{Neepa T. Maitra}
 \affiliation{Department of Physics, Rutgers University, Newark 07102, New Jersey USA}
 \email{Corresponding Author: neepa.maitra@rutgers.edu}
 \date{\today}

\begin{abstract}
Time-dependent density functional theory continues to draw a large number of users in a wide range of fields exploring myriad applications involving electronic spectra and dynamics. Although in principle exact, the predictivity of the calculations is limited by the available approximations for the exchange-correlation functional. In particular, it is known that the exact exchange-correlation functional has memory-dependence, but in practise adiabatic approximations are used which ignore this. Here we review the development of non-adiabatic functional approximations, their impact on calculations, and challenges in developing practical and accurate memory-dependent functionals for general purposes. 
\end{abstract}

\maketitle
\section{Introduction}
\label{sec:intro}
Over the past almost forty years, time-dependent density functional theory (TDDFT) has enabled the calculation of electronic spectra and dynamics in systems that would have been otherwise out of reach to treat quantum-mechanically~\cite{RG84,Carstenbook,TDDFTbook12,M16,LGIDL20,SZYYK21}. 
{\modified While ground-state density functional theory (DFT) is the mainstay of electronic structure, being itself the most widely-used method for materials and molecules as well as the starting point of almost all other treatments of materials, it does not give excitations, or more generally the response to a time-dependent external field whether weak or strong. DFT-flavored methods that do provide excitation energies include $\Delta$SCF~\cite{GL76} which, while originally justified only for the lowest state of a given symmetry, was later shown to have a rigorous basis through the generalized adiabatic connection approach of Ref.~\cite{G99b}; the method however is usually used in a very approximate way with ground-state DFT functionals replacing approximations to the orbital-dependent excited state functionals appearing in the theory.
Ensemble-DFT~\cite{Theophilou1979,GOK88,GOK88b}, recently extensively reviewed in Ref.~\cite{CSRF21}, provides another in-principle exact route to excitation energies, but existing formulations of either ensemble-DFT or $\Delta$SCF do not give access to other response properties such as spectral oscillator strengths. 
 Alternative methods based on approximations to the true wavefunction, or on other reduced quantities such as the one-body Green's function or reduced density-matrices,  require more computational resources.} There are simply no computationally feasible alternatives to TDDFT for some of the applications on complex systems particularly when driven away from their ground states. Some examples over the past five years are described in recent reviews~\cite{LGIDL20,SZYYK21,S21}, and   range from simulations of electronic stopping power~\cite{UAC18}, charge transport in complex molecules~\cite{MFHCGLS22}, across nano-junctions~\cite{JKC22} and in light harvesting systems~\cite{SK18}, attosecond electron dynamics and high-harmonic generation in solids~\cite{MTRD20,S21,Floss18,YNNY18}, laser-driven dynamics in nanogaps of thousand-atom systems~\cite{BCV22}, angle-resolved photo-emission from large clusters~\cite{DPVHSR20}, ultrafast spin transfer~\cite{DESGS18}, Floquet engineering~\cite{Lucchini2022}, and conductivity in a disordered Al system that treated almost 60000 electrons explicitly~\cite{DAGBSC17}.
On the other hand, in the vast majority of cases, TDDFT is applied in the linear response regime, where weak perturbations of the ground-state formulated in the frequency-domain provide excitation spectra and oscillator strengths~\cite{C95,PGG96,GPG00, HC12,AJ13};  the favorable system-size scaling of TDDFT has been further enhanced with the use of embedding methods~\cite{CW04,N07,P13,TBN19,HLPC14,MJW13} or stochastic orbitals~\cite{BNR13,ZLBRN20}. 

The computational efficiency of TDDFT is all the more appreciated considering the climate crisis we face today. Instead of simulating the many-electron interacting TDSE, the Runge-Gross theorem~\cite{RG84,Carstenbook,TDDFTbook12} assures us that we can{\modified, in theory, find the exact time-dependent density and all observables from solving the time-dependent Kohn-Sham (KS) equation}:
\ben
\left( -\nabla^2/2 + v\s(\br,t)\right)\phi_i(\br,t) = i\partial \phi_i(\br,t)/\partial t
\label{eq:tdks}
\een
(in atomic units) where the KS potential is
\ben
v\s(\br,t) = v\ext(\br,t) + v\H[n](\br,t)  + v\xc[n;\Psi_0,\Phi_0](\br,t)
\label{eq:kspot}
\een
Here 
\ben 
v\H[n](\br,t) = \int d^3r' w(\vert\br-\br'\vert)n(\br',t)
\een
 is the Hartree potential, 
 \ben 
 \begin{split}
 n(\br,t) &= N\sum_{\sigma, \sigma_2..\sigma_N}\int d^3r_2..r_N\vert\Psi(\br \sigma, \br_2\sigma_2..\br_N\sigma_N)\vert^2 \\ &= \sum_i^N \vert\phi_i(\br,t)\vert^2
 \end{split}
 \een 
 is the one-body density, and $w(\vert\br-\br'\vert)$ the electron-electron interaction.  The last term in Eq.~(\ref{eq:kspot}) is the time-dependent exchange-correlation (xc) potential, $v\xc[n;\Psi_0,\Phi_0](\br,t)$: a functional of the density, the initial interacting wavefunction $\Psi_0$ and initial choice of KS wavefunction $\Phi_0$. Solving for $N$ single-particle orbitals scales far better with system-size than solving for the correlated wavefunction of $N$ electrons.
 In linear response applications,  a perturbative limit of these equations gives the density response, from which excitation spectra can be extracted; there, instead of $v\xc$, its functional derivative, the xc kernel $f\xc[n_0](\br,\br',t-t')$   is required. What makes this reformulation of many-electron dynamics possible is the Runge-Gross theorem proving the one-to-one mapping between the density and potential for a fixed initial state~\cite{RG84}, and the assumption of non-interacting $v$-representability~\cite{L99}. (We note that the rigorous mathematical foundations of both aspects are somewhat unsettled~\cite{RPL15, FLLS16}). 
 
 {\modified The Runge-Gross theorem guarantees that all observables, beyond just the time-dependent density, can be accessed with knowledge of the corresponding functional of the density and initial KS state. However, the identification of these functionals for observables not directly expressed in terms of the density is a challenging problem that has only been examined in a limited number of studies \cite{WB06,WB07,HKLK09}. In practice, an approximation is made by using the KS wavefunction directly.} {\modified We also note that the theorem holds for the density, but can be generalized to spin-densities, and in practice spin-densities are often used especially when properties related to magnetization are of interest.}

 A key element in these calculations is the xc potential, $v\xc[n;\Psi_0,\Phi_0](\br,t)$, which is unknown and needs to be approximated. 
 Almost all the calculations today use an  adiabatic approximation; that is, a ground-state xc potential evaluated on the instantaneous density. Digging into the theory however reveals that the exact xc potential has {\it memory-dependence}: it depends on the history of the density, $n(\br, t'\le t)$ as well as the initial interacting and KS states, $\Psi_0$ and  $\Phi_0$.   In the linear response regime,  memory-dependence endows the xc kernel with a frequency-dependence. {\modified Ever since} the early days of TDDFT, researchers have been striving to build approximations which include this  memory-dependence. Here, we review these efforts and their successes, reasons for why they are not widely used, and discuss prospects of future developments. Before doing so, we demonstrate, using an exactly-solvable model system, the implications of memory-dependence for both dynamics and excitations, and discuss some exact conditions related to memory dependence.

\section{Significance of memory-dependence}
\label{sec:Hubbard}
The lack of memory-dependence in adiabatic TDDFT has been held responsible for errors in their predictions for many real systems, and sometimes significant failures, e.g. Refs.~\cite{TH00,CZMB04,RN11, RN12, RN12c, HTPI14, WU08, GDRS17,BMVPS18,KVPRC20,QSAC17,GWWZ14,HKLK09,DLYU17}. 
In practise, an adiabatic approximation has two sources of error: one coming from the choice of the ground-state approximation, and one from making the adiabatic approximation itself. In some cases, the spatial non-local property of the xc functional is more important, and is lacking in the commonly used local or semi-local approximations, e.g. for excitons~\cite{BSSR07}, and some charge-transfer excitations~\cite{M17,K17}, and including long-range dependence yields good results even within an adiabatic approximation. But in other cases, memory-dependence is essential in both  real-time non-perturbative dynamics 
and in linear response, as in Refs.~\cite{TH00,CZMB04,RN11, RN12, RN12c, HTPI14, WU08, GDRS17,BMVPS18,KVPRC20,QSAC17,GWWZ14,HKLK09,DLYU17,DPVHSR20}. {\modified In the real-time regime, adiabatic functionals cannot describe resonantly-driven dynamics, or dissipation and relaxation from electron-electron interaction in large systems, for example}.
To isolate the effect of the lack of memory-dependence alone, a useful tool is to consider the adiabatically-exact approximation~\cite{HMB02,TGK08,M16}, which consists of using the exact ground-state (g.s.) approximation:
\ben
v\xc^{\rm A-ex}[n;\Psi_0,\Phi_0](\br,t) = v\xc^{\rm exact \,g.s.}[n(t)](\br)
\een
The `best' adiabatic approximation possible is then to propagate with $v\xc^{\rm A-ex}(\br, t)$, which {\modified would require finding the exact g.s. xc potential at each instant in time}. This is {\modified numerically quite demanding} for all but the simplest systems but very instructive when {\modified carried out}~\cite{TGK08,TK09,RP10,FM14,FM14b}, especially when the resulting density and xc potential can be compared with the exact time-dependent density and xc potential. Finding the latter for a given target density can be generally {\modified achieved} through iteration procedures~\cite{LM18,NRL13,RPL15,FLNM18,JW16,BYW20}, or more simply for two-electron cases with only one doubly-occupied KS orbital, e.g. Refs.~\cite{EFRM12,FERM13,SLWM17,DLFM21,DLM22}.
Sec.~\ref{sec:example} gives an example on a very simple system, comparing propagation with the adiabatically-exact xc potential against  the exact propagation.

In the linear response regime, the KS spectrum is corrected towards that of the true spectrum through the xc kernel which is the density-functional-derivative of the xc potential~\cite{PGG96,C95}. (Sec.~\ref{sec:approx}). The adiabatic approximation yields a frequency-independent kernel, but the frequency-dependence of the exact xc kernel is crucial to capture certain properties: for example, states of double-excitation character in molecules~\cite{TH00,MZCB04}, {\modified corrections to the band gap in semiconductors to which the mere KS gap gives an underestimate~\cite{GVbook,GM12}}, and the excitonic Rydberg series in semiconductors~\cite{GZKT22}. The {\it exact} xc kernel has been found by inversion in a few works~\cite{TK14,EG19,WEG21}, and recently, the general pole structure of the kernel was related to zeros in the density response from counterbalancing behavior of neighbouring oscillatory modes, allowing a parametric reconstruction of the kernel~\cite{GZKT22}.

\subsection{Example: Asymmetric Hubbard Dimer}
\label{sec:example}
To demonstrate the impact of the adiabatic approximation, we consider one of the simplest interacting two-electron systems, the Hubbard dimer: 
\ben
H = -t\sum_{\sigma=\uparrow,\downarrow}(a_{1\sigma}^{\dagger}a_{2\sigma}+a_{2\sigma}^{\dagger}a_{1\sigma})+U\sum_{i}\hat{n}_{i\uparrow}\hat{n}_{i\downarrow}+\sum_{i}v_{i}\hat{n}_{i}
\een
where $t$ is a site-to-site hopping parameter, $U$ is an on-site interaction strength, and $\Delta v = \vert v_{1} - v_{2}\vert$ controls the asymmetry of the two-site system.
This has been thoroughly studied in the ground state, see Ref.~\cite{CFSB15} for a review which also provides a parametrization of the exact ground-state xc potential for this system, tuning the correlation, and examines limiting cases such as the symmetric limit, and the weakly- or strongly-correlated limits. {\modified As emphasized in Ref.~\cite{CFSB15}, while the study of lattice models is insightful and shares  features similar to real-space systems, it is not representative of most TDDFT applications which are done in real-space. Depending on the physical system at hand, the reduction of dimensionality and local nature of the interaction may be valid approximations~\cite{Reiningbook}.} The key parameter in determining the degree of correlation is the ratio of the asymmetry to the on-site interaction, $\Delta v/U$~\cite{CFSB15,CFMB18}; even for large interaction strengths $U$, the system is weakly correlated (meaning, the next lowest energy determinants to the KS ground-state determinant are not ``nearby") if the asymmetry $\Delta v$ is large enough that two electrons essentially sit on the same site in the ground state. Ref. \cite{CFMB18} focuses on the exact features of the exchange-correlation kernel in the same system.

Since  only three singlet states span the Hilbert space, the ground-state Levy constrained search over all wavefunctions can be readily performed numerically given a target density~\cite{B08,LU08,V08b,KSKV10,T11,FFTAKR13,FT12,RP10,SDS13,FM14,FM14b,DSH18,KS18,KKPV13,MRHG14,TR14,RP08,CF12}. This means that, at each time-step, it is straightforward to find the adiabatically-exact xc potential and use it to propagate to the next time-step. In this way, we can directly isolate the effect of making the adiabatic approximation on the dynamics, without any approximation to the ground-state functional. 

\begin{figure}[htbp]
\includegraphics[width=0.45\textwidth]{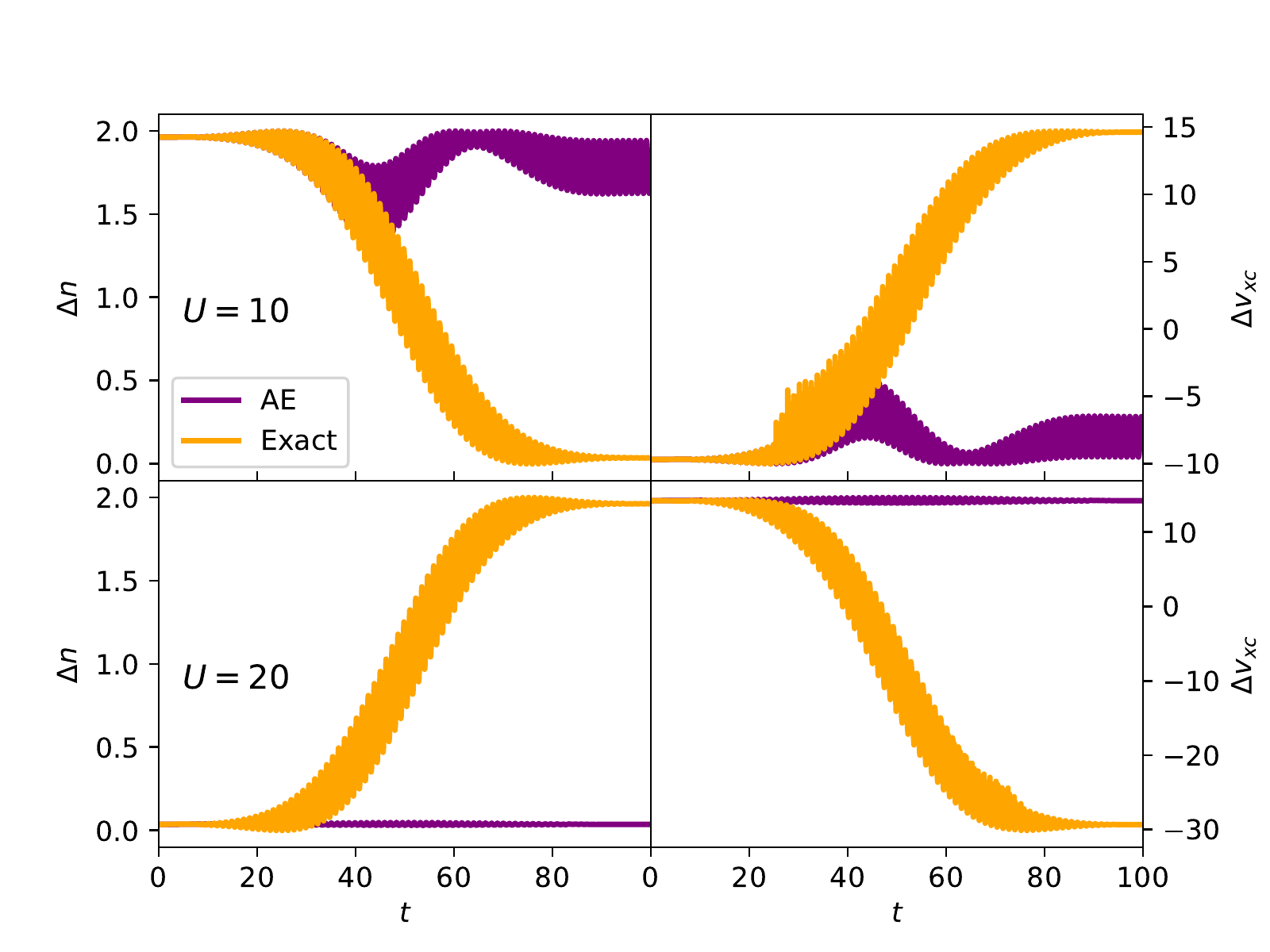} 
\caption{Hubbard dimer driven by a $\pi$-pulse resonant with the lowest excitation: adiabatically-exact (AE) dynamics compared with the exact, in the weak correlation (top panels) and strong correlation (lower panel) cases.  The left-hand side shows the analog of the dipole, the site-occupation difference, $\Delta n = n_1 - n_2$  while the right-hand-side shows the xc potential difference $\Delta v\xc$. 
}
\label{fig:Hubbard}
\end{figure}

\begin{widetext}

\begin{figure}[htbp]
\includegraphics[width=1.\textwidth]{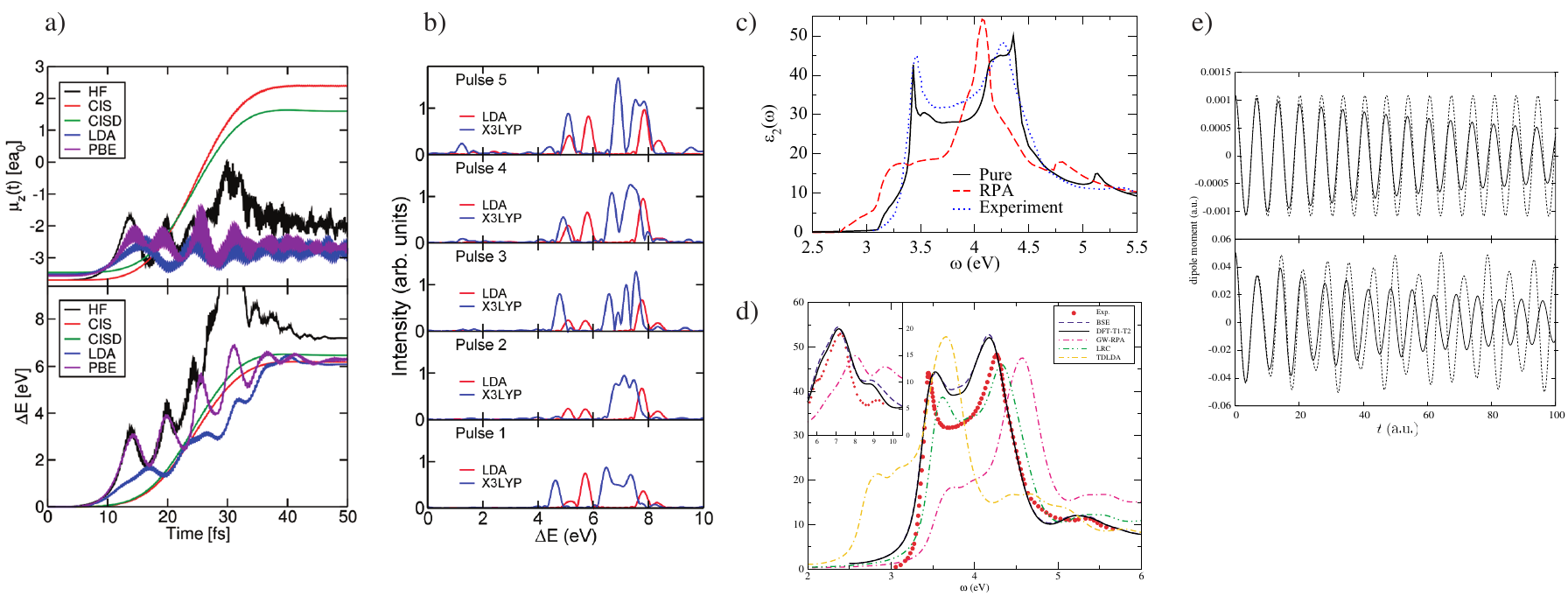} 
\caption{{\modified Examples of errors in predictions  of dynamics in real systems due to the adiabatic approximation:
\newline
a) Charge transfer dynamics in the LiCN molecule driven by a $\pi$-pulse, from Ref.~\cite{RN11}: time evolution of the dipole moment and energies  computed with different methods. The frequency and strength of the pulse are adjusted to represent a resonant single-photon absorption in each case. The adiabatic LDA and PBE functionals are not able to transfer a significant amount of the charge in contrast to the configuration-interaction singles (CIS) and CIS-doubles (CISD) calculations. Analysis on a model system showed that the exact xc potential develops non-adiabatic step and peak features essential in the charge-transfer process~\cite{FERM13}; the lack of these is further related to the spurious pole-shifting~\cite{RN12c,HTPI14,FLSM15}, demonstrated explicitly in the  next panel.
Reprinted (adapted) with permission
from Raghunathan, S., Nest, M., 2011. J. Chem. Theory Comput. 7, 2492-2497. Copyright 2011 American Chemical Society.
\newline
b) Spurious pole-shifting in the electronic structure when LiCN is left in different superposition states after a pulse is applied, from Ref.~\cite{RN12c}: a sequence of short pulses excite the system and the dipole moment between each pulse is recorded whose spectrum is obtained through Fourier transform. The position of the peaks should not change in the exact system, but peak shifting happens due to the adiabatic nature of LDA and X3LYP functionals. (see also Sec.~\ref{sec:Hubbard})
Reprinted (adapted) with permission
from Raghunathan, S., Nest, M., 2012. J. Chem. Theory Comput. 8, 806-809. Copyright 2012 American Chemical Society.
\newline
c) and d) Predictions of the optical response of non-metallic systems underestimate the onset of continuous absorption (i.e. underestimate the gap) as illustrated here from Ref.~\cite{CBR20} in panel c and Ref.~\cite{SOR03} in panel d, via the  imaginary part of the dielectric function of bulk silicon. The onset of absorption is at a too low frequency in both the random phase approximation (RPA) calculation in which the xc kernel is put to zero and the adiabatic LDA (labelled TDLDA), and both miss the excitonic structure evident here in the two-peak shape of the experiment, correctly reproduced by the Bethe-Salpeter equation (BSE) approach~\cite{STP04,BSOSR05}. Although the excitonic feature can be captured by long-range-corrected (LRC) kernels (see also Sec.~\ref{sec:mbpt}), the opening of the gap requires a frequency-dependent kernel~\cite{GM12,TP01} and is related to the derivative discontinuity. Often this is effectively hiding in  a ``scissors shift"~\cite{GGG95} using quasiparticle energies from GW, as done in the ``DFT-T1-T2" of Ref.~\cite{SOR03}, but Ref.~\cite{CBR20} derived a ``Pure" approximation for the discontinuity, obtained entirely from ground-state KS and TDDFT quantities, with an underlying frequency-dependent kernel. 
Reprinted figure with permission from Cavo, S., Berger, J.A., Romaniello, P., 2020. Phys. Rev. B 101, 115109.
 Copyright 2020 by the American Physical Society.
Reprinted figure with permission from Sottile, F., Olevano, V., Reining, L., 2003. Phys. Rev. Lett. 91, 056402. Copyright 2003 by the American Physical Society.
\newline
e) Relaxation dynamics of the dipole moment 
in a doped $\text{GaAs}/\text{Al}_{0.3}\text{Ga}_{0.7}\text{As}$ quantum well, from Ref.~\cite{WU05}: an initial 
 uniform electric field ($0.01$ mV/nm in the top subpanel, and $0.5$ mV/nm lower subpanel) is turned off at $t=0$. Damping is only present when memory is included, via the Vignale-Kohn (VK) functional (Sec.~\ref{sec:TDCDFT}) shown in solid lines, in contrast to the adiabatic LDA in dashed lines. 
Reprinted (figure with permission from Wijewardane, H.O., Ullrich, C.A., 2005. Phys. Rev. Lett. 95, 086401. Copyright 2005 by the American Physical Society.
}}
\label{fig:splash}
\end{figure}

\end{widetext}

Figure~\ref{fig:Hubbard} shows the dynamics driven out of the ground-state by a $\pi$-pulse resonant with the lowest excitation, for parameters $t=1/2$, $\Delta v=-15$, and either $U=10$, or $U=20$ in the strong correlation case. In either case, this excitation has a charge transfer character with respect to the ground state: for $U=10$ the ground state has close to two electrons sitting on the lower site, while the lowest excitation has one electron transferred to the other site, while for $U = 20$ (Mott-Hubbard regime), the ground-state has close to one electron on each state while the lowest excitation has close to two electrons on the lower site. Driving the Hubbard dimer initially in its ground state with a weak $\pi$-pulse at this lowest excitation energy thus shows a large change in the site occupation $\Delta n = n_1 - n_2$ as shown in the left panels of Fig.~\ref{fig:Hubbard}. In both cases, the adiabatically-exact evolution  shows only a small partial charge transfer (particularly small in the strongly correlated case) before returning to oscillate around the ground state density. This failure can be traced to the inability of the adiabatically-exact xc potential to capture dynamical step features~\cite{EFRM12}, in particular, a non-adiabatic step feature associated with charge-transfer~\cite{FERM13,M17,FM14,FM14b}, evident in the oscillations and large change in the xc potentials shown on the right-hand-side; a full discussion of the xc potentials and densities in similar dynamics (using a continuous wave flat envelope driving instead of a $\pi$-pulse) can be found in Refs.~\cite{FM14,FM14b}. 
The dynamics seen here is very similar to that found in real molecules~\cite{RN11} where adiabatic approximations also began to charge transfer and then appeared to give up {\modified(see also Fig.~\ref{fig:splash}) shortly}; the fact that even the adiabatically-exact fails indicates it is an issue with memory-dependence.  

As mentioned above, the problem can be traced to the lack of dynamical steps and peaks in the xc potential, and this appears to especially drastically affect resonant-driving because it leads to spurious pole-shifting: when driven away from its ground-state, the resonant frequencies of a system  predicted by adiabatic TDDFT are {\modified detuned} from the values predicted from linear response of the ground-state~\cite{HTPI14,FLSM15} but excitation energies of a system should not shift with the instantaneous state {\modified (see also Fig.~\ref{fig:splash} shortly)}. The density-dependence of the KS potential results in the response of a general state having spuriously shifted poles, while the exact generalized xc kernel acquires a frequency-dependence that corrects this shift~\cite{FLSM15,LFM16}. Adiabatic TDDFT is trying to predict resonantly-driven dynamics but keeps being driven out of resonance by the density-dependence. 
In fact, one could obtain a larger charge-transfer by instead applying a ``chirped" laser that has a time-dependent frequency that adjusts to the instantaneous resonant frequency of the adiabatic approximation during the driving~\cite{LFM16}. 
In general, the spurious pole-shifting can muddy the interpretation of the underlying dynamics  of molecules when properties of the time-resolved dynamics are measured by a probe; the shifts of peaks in the experimental time-resolved spectrum correspond to different nuclear configurations, while in an adiabatic TDDFT simulation it would be hard to disentangle this from the spurious peak-shifting. 

\begin{figure}[htbp]
\includegraphics[width=0.45\textwidth]{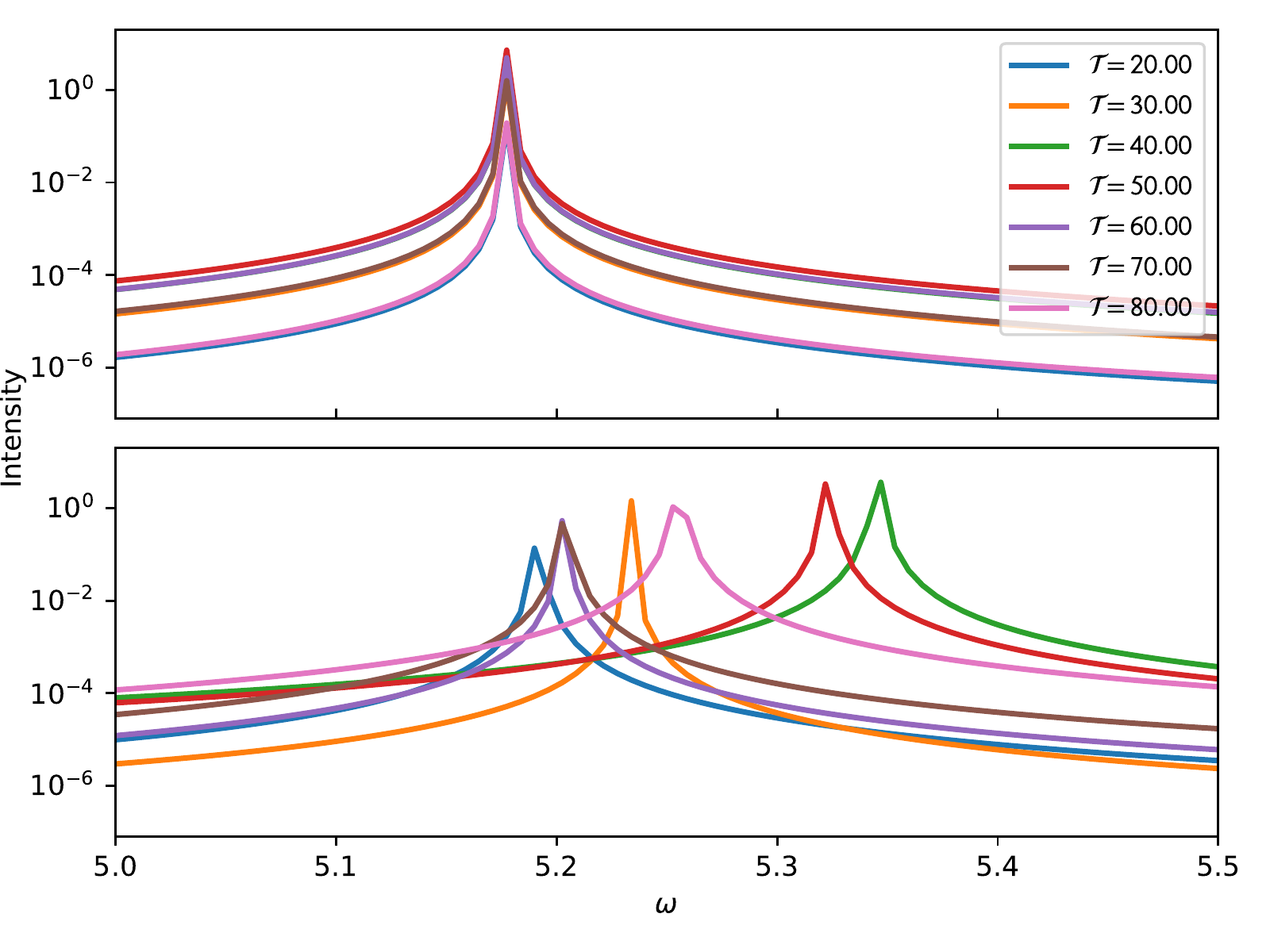} 
\begin{tabular}{| c | c | c || c | c |}
\hline
$\mathcal{T}$ & $\omega^{\rm exact}$ & $\omega^{\rm AE}$ & $\Delta v\ext(\omega^{\rm exact})$ & $\Delta v\ext(\omega^{\rm AE})$ \\
\hline
 $0.00$  & $5.18$ & $5.19$ & $-15.00$ & $-15.01$ \\
 $20.00$ & $5.18$ & $5.19$ & $-15.00$ & $-15.01$ \\
 $30.00$ & $5.18$ & $5.23$ & $-15.00$ & $-15.06$ \\
 $40.00$ & $5.18$ & $5.35$ & $-15.00$ & $-15.18$ \\
 $50.00$ & $5.18$ & $5.32$ & $-15.00$ & $-15.15$ \\
 $60.00$ & $5.18$ & $5.20$ & $-15.00$ & $-15.03$ \\
 $70.00$ & $5.18$ & $5.20$ & $-15.00$ & $-15.03$ \\
 $80.00$ & $5.18$ & $5.25$ & $-15.00$ & $-15.08$ \\
 \hline
\end{tabular}

\caption{Time-resolved spectroscopy in the Hubbard dimer ($U = 10$ case): at the times $\mathcal{T}$ indicated during the dynamics in the top panels of Fig.~\ref{fig:Hubbard}, the field is turned off and the system is probed to measure its absorption spectrum. The top figure shows the exact, the middle shows the predictions from the adiabatically-exact approximation demonstrating the peak-shifting, while the table shows the predictions of the external potential values that would be deduced by comparing the peak position to frequencies predicted from linear response.
}
\label{fig:LM21-eq}
\end{figure}

Ref.~\cite{LM21} argued for the case of using the linear-response TDDFT formulation of Ehrenfest dynamics rather than a real-time formulation due to this problem {\modified when simulating coupled electron-nuclear dynamics}.  This was demonstrated using a model system to compute the errors in predicting the underlying nuclear dynamics when the spectra are calculated in either formulation. 
An analog of this molecular dynamics situation for the Hubbard molecule is demonstrated in Fig.~\ref{fig:LM21-eq}, where the system under the same $\pi$-pulse as in Fig.~\ref{fig:Hubbard} is probed at various times during the pulse by a weak pulse that measures the spectrum at that time. Note that the field is off during the probe measurement, so the exact absorption spectrum should have peaks at the same frequency each time, albeit with different oscillator strength. This condition is not respected by the adiabatically-exact propagation, as evident in the middle panel. The table gives the values of the peak frequencies at the times indicated during the evolution, exact and predicted from the adiabatically-exact evolution,  and  the external potential $v\ext(\omega^{\rm AE})$ whose linear response from the ground-state lies at the dominant peak in the exact and adiabatically-exact calculations. 
At $\mathcal{T}=0$, there is a small difference between the adiabatically-exact and exact transition frequency. Even though the targeted single excitation is far from the doubly-excited state, a small amount of mixing still occurs and its contribution is hidden in the frequency dependent part of the xc kernel that is neglected in the adiabatic propagation.

{\modified Leaving now the Hubbard playground and returning to real systems, Figure~\ref{fig:splash} shows a sampling of results on molecular or solid-state systems where errors due to the adiabatic approximation lead to significant errors in the predicting dynamics or spectra.}

The results here appear to paint a bleak picture for the adiabatic approximation, and yet the TDDFT calculations in e.g. Refs.~\cite{LGIDL20,SZYYK21,DAGBSC17,DESGS18,SK18,YNNY18,Floss18,MTRD20,UAC18,S21,JKC22,MFHCGLS22,DPVHSR20,BCV22}  have been accurate enough to reveal useful information about electron dynamics.
Why this is so, is only partially understood; it is likely that few-electron, few-state systems are the most challenging cases for  TDDFT and that often real systems are complex and large enough that e.g. peaks have significant widths and blur some of the problems discussed here. It is also true that in some applications the dominant effect driving the dynamics is the external potential, and the essential role of the xc potential is to partially counter self-interaction in the Hartree-potential and that a ground-state description of this is adequate. Further, often the observables of interest  in real systems involve averaged quantities (e.g. the dipole moment rather than the spatially-resolved electron density) that can forgive even relatively large local errors in the density~\cite{LM20b}. In situations where the system does not begin in a ground-state, it has been argued that the adiabatic approximation is likely to work best when the KS initial state is chosen to have a similar configuration as that of the true initial state~\cite{EM12,FNRM16,LM20b}, and that if, in the true problem the natural orbital occupation numbers do not significantly evolve even as the natural orbitals themselves may evolve significantly, then the adiabatic approximation can be justified to do a reasonable job even for strongly perturbed dynamics~\cite{LM20b}.  A final consideration is that the adiabatic approximation, by virtue of not having any memory-dependence, in fact satisfies a number of exact conditions that are related to memory (Sec~\ref{sec:excond}).

\subsection{Memory-related exact conditions}
\label{sec:excond}
We note that some insight into the structure of the exact xc potential can be gained from an exact expression resulting from equating the Heisenberg equation of motion for the second time-derivative of the density of the KS system to that of the interacting system~\cite{L99,LFSEM14,FLNM18,LM18,LM20b}:
\ben
v\xc = v\xc^W + v\c^T
\label{eq:exactvxc}
\een
 where the interaction component $v\xc^W$ satisfies
\begin{equation}
  \nabla\cdot(n(\br,t)\nabla v\subxc^W(\br,t)) =  
  \nabla\cdot\Big[
    n(\br,t)\int n\subxc(\br',\br,t)
    \nabla w(\vert\br'-\br\vert)d^3r'
  \Big]\,,
  \label{Eq:theo_vxcw}
\end{equation}
and the kinetic component $v\c^T$ satisfies:
\begin{equation}
  \nabla\cdot(n(\br,t)\nabla v\subc^T(\br,t)) =  
  \nabla\cdot\Big[\mathcal{D}
    (\rho_1(\br',\br,t)-\rho\1s(\br',\br,t))|_{\br'=\br} 
  \Big] \,,
  \label{Eq:theo_vct}
\end{equation}
with $\mathcal{D} = \frac{1}{4}(\nabla'-\nabla)(\nabla^2-\nabla'^2)${\modified, $\nabla=\nabla_\br$ and $\nabla'=\nabla_{\br'}$}. 
Here $n\subxc$ is the time-dependent xc hole defined as
\begin{equation}
  n\subxc(\br,\br') = \rho_{2}(\br,\br';\br,\br')/n(\br') - n(\br)
  \quad,
  \label{Eq:theo_nxc}
\end{equation}
where the {\modified two-body reduced density-matrix} (2RDM) is
\ben
\begin{split}
&\rho_2(\br_{1},\br_{2},\br_{1}',\br_{2}') \\ 
&= N(N-1)\int dr_{3}..dr_{N}
\Psi^*(\br_{1}',\br_{2}',\br_{3},..,\br_{N})
\Psi(\br_{1},\br_{2},\br_{3},..,\br_{N})\,.
\end{split}
\een

The {\modified one-body reduced density matrix} (1RDM) is
\ben
\rho_1(\br,\br') = N\int dr_{2}..dr_{N}
\Psi^*(\br',\br_{2},..,\br_{N})\Psi(\br ,\br_{2},..,\br_{N})\,,
\een
 and $\rho\1s$ is the 1RDM of the KS system. 
 
{\modified Through Eqs.~\ref{eq:exactvxc}--\ref{Eq:theo_vct},} the dependence on the history of the density $n(t'<t)$ and the full interacting and KS initial states $\Psi(0)$ and $\Phi(0)$ is transformed into a more time-local dependence on the KS 1RDM, the true 1RDM, and the xc hole. 
 It has been argued that adiabatic approximations tend to make less error on $v\xc^W$ because the spatial integral appearing there is somewhat forgiving, while $v\c^T$ is responsible for the dominant non-adiabatic effects, including the dynamical steps and peak structures~\cite{FLNM18,LM20b,SLWM17,LSWM18}. Further, the choice of initial KS state has a key 
effect on the size of $v\c^T$, as evident from its dependence on the difference between the interacting and KS 1RDMs; as mentioned earlier, a judicious choice can ease the job of the xc functional approximation.
 Attempts to build a memory-dependent functional based on this decomposition will be discussed in Sec.~\ref{sec:DMxc}.

\begin{figure}[htbp]
\includegraphics[width=0.5\textwidth]{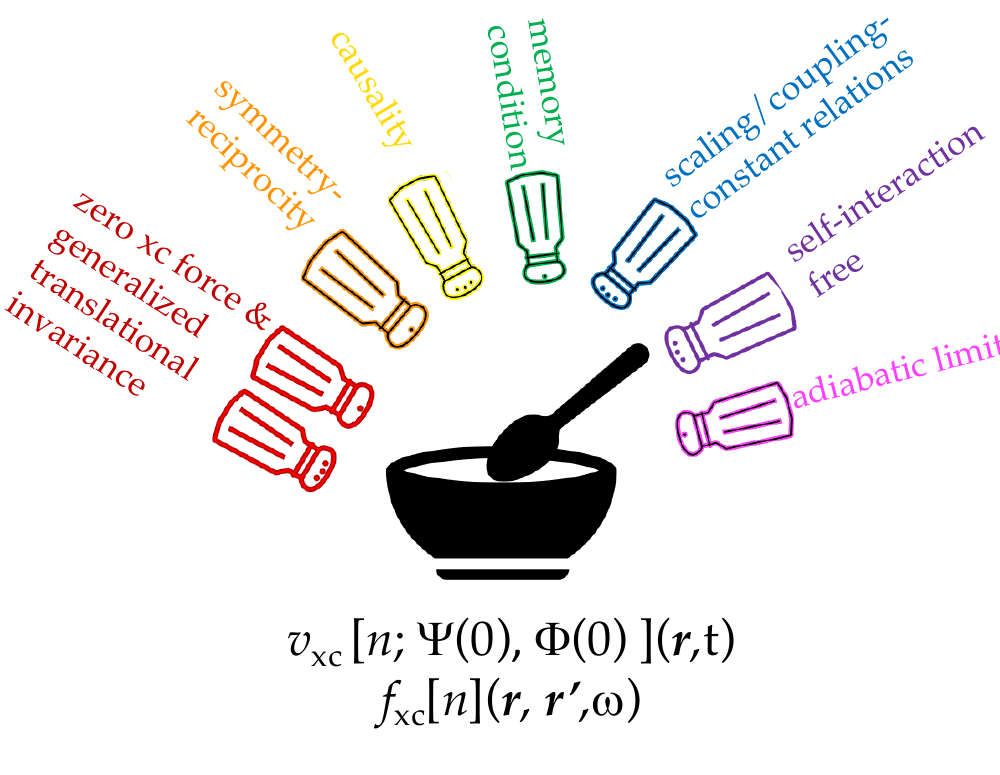} 
\caption{Ingredients for a wholesome functional approximation are some known conditions satisfied by the exact xc potential and xc kernel of TDDFT: the zero force condition and generalized translational invariance~\cite{D94,V95,V95b,GDP96}, symmetry and reciprocity relations~\cite{VK98}, causality (with Kramers-Kronig relations for the kernel)~\cite{L98,WYB12}, memory condition tying together initial-state dependence and history-dependence~\cite{MBW02,DLFM21}, scaling relations~\cite{HPB99}, self-interaction free property~\cite{FLNM18}, and the reduction to the ground-state functionals in the adiabatic limit for systems with a gap~\cite{WYB12}. 
}
\label{fig:sketch}
\end{figure}

Several of the known exact conditions for the ground-state xc functional have analogs in the time-dependent case, e.g. one-electron self-interaction-free conditions ($v\x[n](\br,t) = -v\H(\br,t), v\c[n](\br,t) = 0$), while others are nested in the energy-minimization principle and cannot be {\modified extended} to the time-dependent case (e.g. Lieb-Oxford bound). 
But there are also exact conditions that are inherently associated with the time-dependence of the system, and these typically have implications for memory-dependence of the functional. Here we briefly discuss two of the conditions that have been key ingredients in the development of non-adiabatic approximations, and refer the reader to Refs.~\cite{WYB12,GDP96} and Fig.~\ref{fig:sketch} for discussion of others. {\modified As in the ground-state~\cite{SCAN,PB23},} 
including the ingredients sketched in Fig.~\ref{fig:sketch} in functional approximations leads to increased accuracy and reliability of the TDDFT predictions: Although their importance depends on the type of system of interest (e.g. finite-sized molecule versus extended solid) and the type of dynamics (mere spectra versus far from equilibrium), {\modified they can be responsible for errors in functionals that do not satisfy them, and can lead to} violation of basic quantum principles e.g. unphysical self-excitation~\cite{MKLR07}, spurious pole-shifting in non-equilibrium spectroscopy~\cite{FLSM15,LFM16}.

\subsubsection{Zero Force Theorem}
The zero force theorem~\cite{GDP96,OL90,V95b,V95} (ZFT) ensures that the xc potential does not exert a net force,
  \ben
\int  n(\br,t) \nabla v\xc[n; \Psi_0,\Phi_0](\br,t) d^3 r = 0. 
\label{eq:ZFT}
\een
Since the net force exerted by the Hartree potential vanishes, Eq.~(\ref{eq:ZFT}) ensures that the inter-electron Coulomb interaction does not exert any force on the system (as in Newton's third law of classical mechanics). This also holds in the ground-state, but in the time-dependent case violation of this condition has a particularly severe consequence, leading to numerical instabilities due to the system self-exciting over time~\cite{MKLR07,KB08,SLKSU21}. The linear response limit of Eq.~(\ref{eq:ZFT})  reveals a deep connection between spatial- and time- non-local density-dependence, which we will return to in Sec.~\ref{sec:GK}. 

A related theorem is the net torque theorem~\cite{GDP96,V95b,V95,L01}
\ben
\int n(\br,t) \br \times \nabla v\xc(\br,t) d^3r = \int \br \times \partial_t {\bf j}\xc(\br,t) d^3r
\label{eq:nettorque}
\een
where ${\bf j}\xc = {\bf j} - {\bf j}\s$ is the difference in the current-density of the true system and the KS system~\cite{DLFM21,AV05,MBAG02,SK16,TK09b}.

\subsubsection{Generalized translational invariance } 
Translational invariance requires the wavefunction in an accelerated, or ``boosted", frame to transform as,
\ben
|\Psi^\bb (\br_1...\br_N,t) \rangle
=  \prod^N_{j=1}e^{-i \br_j \cdot\dot{\bb}(t)}|\Psi (\br_1+\bb(t)...\br_N+\bb(t),t)\rangle
\label{eq:boostedwf}
\een
where $\bb(t)$ is the position of the accelerated observer and   
$\bb(0)=\dot{\bb}(0)=0$ such that the accelerated and inertial systems coincide at the initial time. 
The boosted density transports rigidly, 
\ben
n^\bb(\br,t)=n(\br+\bb(t),t).
\label{eq:rigiddens}
\een
 Ref.~\cite{V95} proved that 
in order to fulfill Eq.~\ref{eq:rigiddens} the xc potential must transform as
\ben
  v\xc^\bb[n; \Psi(0), \Phi(0)](\br,t) = v\xc[n;\Psi(0), \Phi(0)](\br+\bb(t),t)\,.
  \label{eq:GTI}
\een
In fact, a $v\xc$ that fulfills Eq.~(\ref{eq:GTI}) automatically fulfills the ZFT Eq.~(\ref{eq:ZFT}) \cite{V95}. The Generalized Translational Invariance (GTI)  and ZFT are closely related, particularly in the linear response regime~\cite{VK98}.
 A special case of Eq.~(\ref{eq:GTI}) is the harmonic potential theorem (HPT), which states that for a system 
 confined by a harmonic potential and subject to  a uniform time-dependent electric field, the density transforms rigidly following
 Eq.(\ref{eq:rigiddens}) where $\bf{b}(t)$ is the position of the center of mass ~\cite{D94,V95,GDP96,Carstenbook}.

\section{Non-adiabatic approximations in TDDFT}
\label{sec:approx}
From soon after the birth of TDDFT to today, non-adiabatic approximations have been derived and tested. 
We review these, in roughly chronological order, below. The earliest ones focussed on the linear response regime, where  instead of needing an approximation for the full xc potential $v\xc[n; \Psi_0,\Phi_0](\br,t)$, an approximation for the xc kernel $f\xc[n_0](\br,\br',t - t') = \frac{\delta v\xc(\br,t)}{\delta n(\br',t')}$, or its frequency Fourier transform is needed~\cite{GK85,PGG96,C95}. The TDDFT linear response formalism is based on the density-density response function, $\chi(\br,\br',\omega)$ which describes the  linear density response of the system at frequency $\omega$ to a perturbation $\delta v$: $\delta n(\br,\omega) = \int d \br' \chi(\br,\br',\omega)\delta v(\br',\omega)$. In TDDFT, 
\ben
\begin{split}
\chi(\br,\br',\omega) &= \chi\s(\br,\br',\omega) \\ & + \int dr_1 dr_2 \chi\s(\br,\br_1,\omega) f\Hxc(\br_1,\br_2,\omega) \chi(\br_2,\br',\omega)
\end{split}
\een
 where 
$\chi\s$ is the KS density-response function, giving the density-response to a perturbation of the KS potential $v\s$, and  $f\Hxc(\br_1,\br_2,\omega) = \frac{1}{\vert\br_1 - \br_2\vert} + f\xc[n_0](\br_1,\br_2,\omega)$. Being evaluated at the ground-state density eliminates the initial-state dependence,  due to the Hohenberg-Kohn theorem of ground-state DFT~\cite{HK64}, and memory-dependence corresponds to frequency-dependence of the xc kernel, since $\frac{\delta v\xc(\br,t)}{\delta n(\br',t')}$ is not merely proportional to a delta-function in the time-difference which gives a constant in the frequency Fourier transform.  The instantaneous dependence of the xc potential with respect to the density in the adiabatic approximation yields a frequency-independent kernel. 

\subsection{Finite-frequency LDA}
\label{sec:GK}
{\modified The Gross-Kohn (GK) approximation~\cite{GK85,GK90} is a straightforward first attempt to extend the LDA into the dynamical regime, retaining the spatially-local dependence on the density while introducing time-nonlocal dependence via the finite-frequency response a uniform electron gas.}
That is, the xc kernel is approximated by
\ben
f\xc^{\rm GK}[n_0](\br,\br',\omega) = \delta(\br - \br') f\xc^{\rm unif}[n_0(\br)](q = 0, \omega)
\label{eq:GK}
\een
where $f\xc^{\rm unif}[n](q, \omega) = \int f\xc^{\rm unif}[n](\br, \br', \omega) e^{i {\bf q}\cdot (\br - \br')}d\br'$ is the xc kernel of a uniform electron gas of ground-state density $n$ at wavevector ${\bf q}$ and frequency $\omega$. 
The time-nonlocal dependence of the GK approximation is explicit when considering the response xc potential in the linear response regime corresponding to Eq.~(\ref{eq:GK}):
$
v\xc^{\rm GK}[n](\br,t) = \int f\xc^{\rm unif}[n_0(\br)](t-t')\delta n(\br,t') dt'
$.

The uniform electron gas xc kernel, $f\xc^{\rm unif}[n](q, \omega)$ is known exactly {\modified for a range of densities} in some limits, e.g. in the long-wavelength limit $q = 0, \omega\neq 0$ in Refs.~\cite{GK85,IG87}, and the static limit $q \neq 0, \omega = 0$ in Refs.~\cite{CP07,RA94} where parameterizations based on quantum Monte Carlo  have been developed in Ref.~\cite{CSOP98}.  {\modified An early interpolation between the two limits was given in Ref.~\cite{Dabrowski86} focussing on the metallic density range. More recently, Ref.~\cite{RNPP20} developed an interpolation over a wide range of densities while incorporating first-principles constraints, with Ref.~\cite{KNRBP22} building upon this, embracing even lower densities and more exact conditions. }
  {\modified We note that taking the limit $\omega \to 0$ first then $q \to 0$ of the kernel reduces to the adiabatic local density approximation (ALDA). The order in which the limits $q \to 0$ and $\omega \to 0$ are taken is crucial, as the outcome depends on their order~\cite{GVbook,VK98}.}
 It should also be noted that a fundamental difference between the uniform gas kernel and that of inhomogenous systems is the long-wavelength finite-frequency behavior as $q \to 0$, where $f\xc^{\rm unif}(q \to 0, \omega)$ tends to a finite constant, while for {\modified non-metallic} systems, the kernel diverges as $\alpha(\omega)/q^2$ which has important consequences for the optical response of solids~\cite{ORR02,BSSR07,TDDFTbook12} (see also Sec.~\ref{sec:mbpt}).  {\modified Finally, we note that Ref.~\cite{PGR18} tabulated $f\xc^{\rm unif}[n](q, \omega)$ for a wide range of wave-vectors and frequencies through a correlated equations of motion approach including single particle-hole and two-particle two-hole excitations, within a correlated basis functions formalism for computing matrix elements. This revealed a double-plasmon excitation for which such a non-adiabatic kernel is essential (c.f. Sec.~\ref{sec:specific}), and Ref.~\cite{PGR18} further showed the possibility of using these results to obtain such features in spectra of inhomogeneous systems.}

{\modified Returning to the GK approximation},  Eq.~(\ref{eq:GK}) can be viewed as a ``double LDA", in that both the ground-state density of the system $n_0(\br)$ varies slowly enough in space that the density-functional argument can be replaced by the local density considered as part of a homogeneous system, and also that the response of the system varies slowly enough that only the zero wavevector component is used.

 A question is whether an approximation may still be reasonably accurate well beyond the situation for which it was derived, as is the case for the ground-state LDA. In the ground-state case, a key reason often given for why LDA gives useful results for non-uniform densities of molecules and solids is its satisfaction of exact conditions~\cite{B12,PK03} such as sum-rules on the xc hole. 
Unfortunately, the GK approximation violates several of the exact conditions discussed in Sec.~\ref{sec:excond} that are important for the time-dependent problem. Refs.~\cite{D94,V95} pointed out that it violates the HPT: instead of yielding a rigid sloshing of the density in a uniform field-driven harmonic well with a center that follows the classical center of mass motion,  GK results in a density-dependent shift in the frequency of this motion, and a damping of the oscillations. The problem is that a potential $v\xc(\br,t)$ that sees only the density at that point $\br$ over time cannot tell whether the change over time is from a sloshing motion or a compression/expansion motion. 

In fact, GK's violation of the GTI condition, of which HPT is a special case, was shown in Refs.~\cite{V95,V95b} to imply that the xc kernel for a non-uniform system at finite frequency has a density-dependence that is long-range in space  such that a local-density or gradient expansion approximation simply does not exist. The ZFT also demonstrates this: taking the linear-response limit of Eq.~(\ref{eq:ZFT}) by writing $n(\br,\omega) = n_0(\br) + n_1(\br,\omega)$ and $v\xc(\br,\omega) = v\xc[n_0](\br) + \int f\xc(\br,\br',\omega) n_1(\br',\omega) d^3r'$, one arrives at
\ben
\int f\xc[n_0](\br,\br',\omega)\nabla n_0(\br') d^3 r' = \nabla v\xc[n_0](\br)
\label{eq:GK-contradiction}
\een
 Inputting the GK kernel Eq.~\ref{eq:GK} yields $\nabla n_0(\br) f\xc^{\rm unif}[n_0(\br)](q = 0, \omega)$ on the left-hand-side, which is a clearly frequency-dependent quantity, quite incompatible with the right-hand-side which is frequency-independent! The argument can be generalized to show that even a short-ranged xc kernel violates the condition when applied to slowly-varying ground-state densities, and that $\int f\xc(\br,\br',\omega) d^3r'$ must diverge. 
In fact,  Eq.~(\ref{eq:GK-contradiction}) implies that spatial and time non-local density-dependence are intimately related in the exact xc kernel, since the spatial integral of the frequency-dependence must yield a frequency-independent quantity. 
Thus the ZFT, which is a seemingly natural statement embodying simply a Newton's third law type of physics leads to quite an extraordinary result:  time non-local density-dependence implies spatially non-local dependence, and that a local-density approximation, or gradient expansion, with memory simply does not exist~\cite{Vignalechap,VK98}. Since this is is true even in the limit of slowly-varying densities, the effect has been dubbed ``ultra-nonlocality". 

The effect of the violation of the GTI and ZFT by the GK approximation on dynamics was shown to yield unphysical behavior on dynamics in sodium clusters in Ref.~\cite{KB08}, including the appearance of spurious  low-frequency modes and instabilities. Ref.~\cite{KB08} also presented a way to directly impose these conditions through constraints on the potentials, applicable in principle to any xc kernel or potential approximation. 

\subsection{Time-Dependent Current-Density Functional Theory}
\label{sec:TDCDFT}
While Ref.~\cite{D94} pointed out GK's violation of the HPT theorem, it also suggested an avenue for a remedy: to separate out the translational motion of the density from that of the compression/expansion and to use the finite-frequency uniform-gas kernel only for the latter. Ref.~\cite{DBG97} further developed this notion that memory resides with a ``fluid element" and that, although there is no spatially-local time-nonlocal description in terms of $n(\br,t)$, we can search for such a description in terms of $n(\bR(t'))$ where $\bR(t') = \bR(t' | \br t)$ is the position of the fluid element at time $t'$ which at $t$ is at position $\br$. In this way, the current-density  naturally enters the picture because this is what dictates the trajectory $\bR(t)$, through $\partial_{t'}\bR(t'|\br,t)  = {\bf J}(\bR,t')/n(\bR,t')$ with the boundary condition $\bR(t |\br,t) = \br$. The Dobson-B\"unner-Gross functional thus applies the GK functional in a frame that moves with the local velocity $ {\bf J}(\bR,t')/n(\bR,t')$, resulting in a functional that satisfies the HPT and GTI~\cite{DBG97}.

At around the same time (in fact the ``received" date is earlier for Ref.~\cite{DBG97} than for Ref.~\cite{VK96} although the ``published" date is after), the  Vignale-Kohn current-density functional was developed, which elevated the current-density from simply assisting to actually being the basic variable of the functional~\cite{VK96,VK98,VUC97}. Time-dependent current-density functional theory (TDCDFT) is based on the one-to-one mapping between the current-density and the vector potential acting on the system, for a given initial state, which had been proven earlier in Ref.~\cite{GD88,V04}. 
The KS equation has the form
\ben
\left(\frac{(-i\nabla + {\bf a}\s(\br,t))^2}{2} + v\s(\br,t)\right)\phi_i(\br,t) = i \partial_t{\phi_i(\br,t)}
\een
where there is a gauge-freedom between the longitudinal part of the KS vector potential and scalar potential, e.g.  putting all the time-dependent external fields and xc fields into the vector potential would yield ${\bf a}\s(\br,t) = {\bf a}\ext(\br,t) + {\bf a}\xc(\br,t)$, $v\s(\br,t) = v_{{\sss ext},0}(\br) + v\H(\br,t) + v_{{\sss XC},0}(\br)$, but other gauge choices can be made.
Even if only a scalar potential is applied to an interacting system, the resulting current-density is typically only reproducible by a non-interacting system with a vector potential; the TDDFT KS current-density usually differs from the physical current-density by a rotational component, even when the exact xc functional is used~\cite{AV05,MBAG02,TK09b,SK16,DLFM21}. 

A key motivation for using TDCDFT is that the current-density at a given point in space contains spatially non-local density-dependence, which can be seen from inverting the continuity equation: $\nabla\cdot{\bf j}(\br,t) = -\partial_t n(\br,t) \rightarrow {\bf j}_L(\br,t) = \int \partial_t n(\br',t)\nabla_r \frac{1}{4\pi \vert \br - \br'\vert} d^3 r'$, which implies that spatially local functionals of the current-density have spatially non-local density-dependence. In fact a local  approximation in terms of the current-density does exist, and, when considered through the linear current response of a slowly spatially-varying electron gas, forms the basis of the
Vignale-Kohn (VK) approximation~\cite{VK96,VUC97,VK98,Vignalechap}. The central role is played by the tensorial xc kernel which is the functional derivative of the xc vector potential with respect to the current-density, for the ground-state of a slowly-varying periodically modulated electron gas.  This functional is constructed in the linear response regime, built up from both longitudinal and transverse responses of the uniform electron gas~\cite{QV02, NCT98,CNT97,CV99} together with the imposition of  several exact conditions: the zero force and zero torque identities (Sec.~\ref{sec:excond}), the Ward identity and symmetry/reciprocity relations~\cite{VK98}.  The resulting functional for the linear response xc vector potential  takes the following form~\cite{VK96,VUC97,VK98}:
\ben
\frac{-i\omega}{c}a_{{\sss XC}1, i} = -\nabla_i v_{{\sss XC}1}^{\rm ALDA}(\br) + \frac{1}{n_0(\br)}\sum_j \frac{\partial \sigma_{{\sss XC},ij}(\br, \omega)}{\partial r_j}
\een
where 
\ben
\sigma_{{\sss XC},ij} = \tilde\eta\xc\left(\frac{\partial u_i}{\partial r_j} + \frac{\partial u_j}{\partial r_i} - \frac{2}{3}\nabla\cdot{\bf u} \delta_{ij} \right) + \tilde\xi\xc \nabla\cdot {\bf u} \delta_{ij}
\een
where ${\bf u} = {\bf j}(\br)/n_0(\br)$ is the velocity field, and $\tilde\eta$ and $\tilde\xi\xc$ are complex viscosity coefficients, expressible in terms of the longitudinal and transverse response kernels of the uniform electron gas, and functions of the frequency and ground-state density~\cite{VK96,VUC97,VK98}. Although the original form looked more complicated, it was shown to be equivalent to the form above in Ref.~\cite{VUC97}, giving a physical interpretation in terms of a Navier-Stokes form for the current-density where a hydrodynamical viscoelastic stress tensor has complex viscosity coefficients of the electron liquid. 
The VK approximation becomes exact in the limit that the length scale of the variations of the ground-state density $q^{-1}$ and perturbing potential $k^{-1}$ are such that $k, q << \omega/v_F, k_F$ where $k_F, v_F$ are the local Fermi momentum and velocity . Therefore, this theory is
applicable to the study of high frequency phenomena, but due to its satisfaction of some exact conditions and spatially-non-local density-dependence, it has also been successfully applied in the static regime where long-range effects are important (more shortly). 
It has also been extended to the non-linear regime in Ref.~\cite{VUC97}. Building on the VK approach, Ref.~\cite{TV06,TVT07} derived a GGA and meta-GGA to move beyond the slowly spatially-varying assumption.

The Vignale-Kohn approximation has memory-dependence and spatially nonlocal
density-dependence (local in current). 
Due to these features, it has successfully
predicted linewidths of collective modes in two-dimensional quantum strips and quantum wells~\cite{UV98,UV98b,UV01} absent in LDA or GGA, time-resolved dissipation from electron-electron interaction in large or periodic systems~\cite{AV06,U06} missed in LDA or GGA or with any adiabatic functional, stopping power in metals~\cite{NPTVC07}, spin-Coulomb drag~\cite{AU06},  and static polarizabilities in long polymer chains~\cite{FBLB02,FBLB03} (routinely
underestimated by adiabatic LDA or GGA). However, it has been shown to generate unphysical
damping of excitations and dissipation in finite systems~\cite{UB04}, {\modified does not provide a significant correction to the band-gap}, and does not work well for the optical response of semiconductors unless either an empirical factor is used~\cite{BKBL01,BBL07} or it is combined with other non-empirical polarization functionals where it corrects for bound exciton widths in insulators and semiconductors and Drude tails in metals~\cite{B15}. Even if only
applied to metallic extended systems, some caution should be applied since the longitudinal and transverse electron gas response
functions entering the Vignale-Kohn functional have some uncertainty for general
frequencies and wavevectors~\cite{QV02,CNT97}. 
The non-linear extension
of the Vignale-Kohn approximation has also been used to study decoherence
and energy relaxation of charge-density oscillations in quantum wells~\cite{WU05}. A different nonlinear non-adiabatic functional based on Landau Fermi liquid theory was presented in Ref.~\cite{TP03} and was the precurser of deformation functional theory which will be shortly discussed. 

\subsection{TDCDFT via an action functional}
TDCDFT was also the framework for a general formulation where memory is included through an action functional defined on the Keldysh contour in order to preserve causality~\cite{KB06}: $v\xc$ should depend only on the past-density, not the future, which means its functional derivative $f\xc[n](\br,t-t')$ should be zero for $t >t'$~\cite{L98}
and this property would be violated if $v\xc$ was the functional derivative of an action  defined in physical time rather than on the Keldysh contour.   
 (We note that some caution is needed when using the Keldysh contour in TDDFT:  the Runge-Gross one-to-one mapping on the Keldysh contour has not yet been proven~\cite{L01} but the contour can be still be used rigorously in variational formulations e.g. as demonstrated in Ref.~\cite{T07}). 
Regarding a real-time resolution of the causality paradox, we refer the reader to Ref.~\cite{V08}.

The use of
  Lagrangian coordinates in the action functional resulted in xc potentials that again preserve the GTI and ZFT~\cite{KB04}; the Lagrangian description arises naturally when thinking of the convective fluid element motion in TDCDFT. The same authors developed a computationally simpler approach that avoids Lagrangian frames, and instead constructs a family of translationally invariant actions on the Keldysh contour which automatically satisfy GTI and ZFT, and they derive a memory-correction to ALDA built using the uniform electron gas kernel of Ref.~\cite{GK85,IG87} in this framework~\cite{KB06}. The effect of memory in this approximation was highlighted in creating viscous effects in both the linear and non-linear regime in plasmon dynamics and absorption in spherical jellium gold clusters.

\subsection{Time-dependent deformation functional theory}
 When TDCDFT is recast in the Lagrangian frame, the natural spatial coordinate to use at time $t$ becomes the initial point ${\bf \xi} = {\bf \xi}(\br,t)$ of the trajectory which  at time $t$ is at position $\br$. Ref.~\cite{T05,T05b,T07,TokatlyChap} showed that by separating the convective motion of the ``electron fluid elements" from their relative motion, the many-body effects  are contained in an xc stress tensor which depends on the the time-dependent metric tensor of the $\br \to \xi$ transformation; since this tensor corresponds to Green's deformation tensor of classical elasticity theory, the approach is called time-dependent deformation functional theory (TDdefFT). 
This framework is fully non-linear from the start and offers a distinct starting point for approximations. The local deformation approximation is based on uniform time-dependent deformations of the uniform gas, giving a stress tensor with spatially-local but time-non-local dependence on the metric tensor; this does not violate the ZFT and GTI since the convective non-locality is treated exactly through its dependence on the Lagrangian coordinate $\xi(\br,t)$.  Ref~\cite{UT06} compared TDdefFT and TDCDFT in both linear and nonlinear models of charge-density oscillations. In the limit of small deformations, the local approximation in TDdefFT reduces to the VK TDCDFT approximation~\cite{T05b}. Another limit is an elastic one, valid for very fast variations of the deformation tensor. This is spatially-nonlocal and related to the ``antiadiabatic" limit of the xc kernel~\cite{NTPV10}. 
In ``quantum continuum mechanics"~\cite{TGVT09,GTVT10,GJTD12}, the hydrodynamic picture has been applied directly to the many-body system without being propped up by a KS system.

\subsection{Orbital functionals}
\label{sec:orb}
Hydrodynamic methods that are based on xc effects of uniform or slowly-varying electron gases are problematic for finite systems, since they introduce spurious dissipation. A consequence in the linear response regime is that the predicted excitation energies of atoms and molecules attain an unphysical lifetime~\cite{UB04}.  
A different direction to incorporate memory and also spatial nonlocality is to develop explicit functionals of the KS orbitals $v\xc[\{\phi_i\}](\br, t)$. Since each orbital itself depends on the density in a spatially and time non-local way, an explicit local functional of the orbital is an implicit non-local functional of the density.  Some common orbital functionals are meta-GGAs, which are have semi-local spatial dependence on the orbitals, hybrids incorporating a fraction of exact exchange, and self-interaction corrected LDA; the latter two have non-local spatial dependence from Coulomb integrals between orbitals and so are computationally more expensive. The spatial non-locality enables various properties in TDDFT to be better reproduced for reasons unrelated to memory, e.g. capturing the $-1/r$ asymptotic decay of $v\xc$ which reclaims the Rydberg series of excitations in the bound spectrum that are otherwise lost in the continuum~\cite{WMB03}, particle-number discontinuities in the xc kernel that are important in capturing charge-transfer excitations~\cite{HG12}, and ionization~\cite{TGK08}. One advantage of these approaches is that self-interaction error is more easily dealt with, unlike in the hydrodynamic approaches in Sec.~\ref{sec:TDCDFT}.

There are two fundamentally different but formally rigorous ways in which to treat orbital functionals. In one, as in KS theory with explicit density-functionals, the xc potential is the same function for all orbitals, and, when the approximation enters through an xc action $A\xc[\{\phi_i\}]$, possibly on the Keldysh contour~\cite{L98}, $v\xc[\{\phi_i\}](\br,t) = \frac{\delta A\xc[\{\phi_i\}]}{\delta n(\br,t)}$ is obtained through the time-dependent orbital effective potential equations (TDOEP)~\cite{UGG95}. 
Exact exchange has been computed in this way~\cite{G97,G98,HB08,HB09,HIG09} primarily in the linear response regime for the computation of excitation energies. For example, for long-range charge-transfer excitations between closed-shell fragments, the importance of derivative discontinuities of the xc kernel with respect to particle number was shown to play a key role, and the exact exchange kernel, due to its orbital dependence captures these, and yields the correct asymptotic behavior of these excitation energies when used in its fully non-adiabatic form~\cite{HG12,HG13,M22}. 
A finite derivative discontinuity is related to correcting the self-interaction from the Hartree functional~\cite{P90}, and
self-interaction corrected LDA has also been applied within TDOEP~\cite{HKK12}. 

The second way to treat orbital functionals resulting from an action functional (or in the ground-state case an energy functional), is through the so-called generalized KS approach~\cite{SGVML96,GL97b,GNGK20}, resulting in orbital-specific xc potentials. This avoids having to solve the numerically challenging TDOEP equation, and is the most common treatment of hybrid functionals where a fraction of Hartree-Fock exchange is mixed in. Hybrid functionals have some advantage in the ground-state in that the delocalization error of semi-local functionals is partially compensated by the localization error of Hartree-Fock~\cite{YCM12}, but perhaps most interesting for TDDFT is that the meaning of the predicted excitations is fundamentally different for the pure KS treatment than for the generalized KS treatment. In the former, excitations are those of the neutral system, while in the latter, they have a character in between neutral and addition (or affinity) energies, depending on the amount of Hartree-Fock mixed in. This has proved particularly useful for charge-transfer excitation energies when range-separated hybrids are used~\cite{BLS10,KSRB12,SKB09,KB14,KKK13,K17b,M22,M17}.

However, coming back to the theme of non-adiabaticity, without frequency-dependence, a general treatment of charge-transfer excitations in response and charge-transfer dynamics in real-time, are out of reach for hybrid functionals whether treated in generalized or pure KS. This can be most easily seen with the two-electron example, where ground-state exact exchange and Hartree-Fock exchange coincide, equaling half the Hartree potential: a frequency-independent xc kernel cannot properly describe excitations when there are KS determinants lying near the ground-state(also Sec.~\ref{sec:specific}) as in the case of a stretched heteroatomic diatomic molecule~\cite{M05c},
and the non-adiabatic dynamical steps and peaks mentioned in Sec.~\ref{sec:Hubbard} necessary to achieve the transfer of an electron in real time are missing~\cite{FERM13,M17}. 

There have been far fewer applications of orbital functionals in the nonlinear regime, largely because of the computational expense~\cite{WU08,HKK12,LHRC17,LHRC18}. Applying the KLI approximation~\cite{KLI92} simplifies the TDOEP calculation~\cite{UGG95,URS98,HTC13} but becomes problematic because  of its violation of the ZFT which can result in unphysical self-excitation~\cite{MKLR07}.

\subsection{Bootstrapping Many-Body Perturbation Theory}
\label{sec:mbpt}
Noting that the density is the diagonal of the one-body Green's function in the equal-time limit, an exact integral expression for the time-dependent xc potential can be found through the Dyson equation for the interacting Green's function when referred to the KS Green's function~\cite{L96}. Taking place on a Keldysh pseudotime contour, this is the time-dependent generalization of the Sham-Schl\"uter approach~\cite{SS83, S85}, and connects $v\xc(\br,t)$ to the two-point self-energy $\Sigma\xc(\br,t; \br',t')$ of many-body theory. Approximations derived from many-body diagrammatic expansions can thus be transformed into equivalent approximations for the TDDFT xc potential. These often result in orbital functionals with implicit density-dependence (Sec.~\ref{sec:orb}). The first order in perturbation theory is the Hartree-Fock approximation for $\Sigma\xc$ which, through the time-dependent Sham-Schl\"uter equation yields the time-dependent exact exchange potential, while higher order perturbation theory introduces correlation. A diagrammatic expansion of the equation using the KS Green's functions as the bare propagators showed that the spatial nonlocality of the xc kernel is strongly frequency-dependent, and related the long-ranged divergence at frequencies of excitation energies to the discontinuity of the xc potential~\cite{TP01}. 

To ensure that the selection of diagrams respects fundamental conservation laws such as particle number, energy, momentum, and angular momentum conservations,  instead a variational formalism similar to many-body perturbation theory~\cite{Baym62,BK61} has been developed~\cite{BDLS05}.  One defines a universal functional of the Green's function for the non-classical electron interaction, of which the self-energy is the functional derivative. The total action  (or energy, in the ground-state case) as a functional of the Green's function then has its stationary point at the Green's function that solves the Dyson equation with the consistent self-energy, and restricting the functional domain to that of Green's functions arising from non-interacting Schr\"odinger equations with local multiplicative potentials, results in a procedure to obtain TDDFT approximations~\cite{BDLS05}. The linearized Sham-Schl\"uter equation, in which the Green's function is replaced by the KS Green's function everywhere, can be derived from such an approach. 

Other work where TDDFT approximations have been derived from connections with many-body perturbation theory have been within the linear response regime, and largely focussed on non-metallic extended systems, using Bethe-Salpeter formalism~\cite{ORR02,BSSR07}. The latter involves the four-point reducible polarizability which needs to be contracted to the two-point density-response function to relate to TDDFT; we refer the reader to Ref.~\cite{BSSR07} for a review on different ways in which this has been {\modified achieved} by different groups. In particular, for semiconductors, the exact xc kernel can be considered as a sum of two terms~\cite{STP04,BSOSR05}: one that changes the KS band gap to the larger quasiparticle one, and the other that accounts for the electron-hole interaction responsible for excitonic effects. The first term is usually just accounted for by using the quasiparticle gap, without explicitly finding the kernel that opens the gap; non-adiabaticity is needed for the kernel to achieve this (see an argument by Vignale described in Ref.~\cite{GM12}). For the second term, an expression was derived involving contractions over the quasiparticle polarizability, one-body interacting Green's function and screened Coulomb interaction, and various approximations considered and tested~\cite{BSSR07,RORO02,SOR03,ASM03,MSR03,STP04}. From the point of view of non-adiabaticity, frequency-dependence was shown to arise out of the contractions over spatial indices, even when the many-body quantities being integrated over are frequency-independent~\cite{GORT07}.
The calculation of the two particle matrix elements is expensive, and instead the separation has been exploited to define a two-part procedure to calculate accurate optical spectra of semiconductors and insulators from first-principles without any empirical parameters and without any calculations or input needed outside TD(C)DFT~\cite{CBR20}: first, a modified KS response function is defined through an implicit kernel that accounts for the derivative discontinuity giving rise to gap-opening and second, a polarization functional from TDCDFT is applied for the part of the kernel that captures excitonic effects~\cite{B15}. While spatially long-ranged behavior is essential for the latter, time-nonlocality (frequency-dependence) is implicitly contained in the former to open the gap.

\subsection{Density-matrix coupled approximations}
\label{sec:DMxc}
Another approach focusses on building approximations to the one-body and two-body reduced density-matrix (1RDM, 2RDM) that appear in an exact expression for the xc potential, Eq.~(\ref{eq:exactvxc}).
If the interacting 2RDM  could be somehow modeled, it would provide the two ingredients in the exact $v\xc$ of Eq.~(\ref{eq:exactvxc}) that are not directly accessible from the TDKS evolution, $\rho_1$ and $n\xc$. A particular class of such approximations, denoted ``density-matrix coupled approximations", couples the TDKS equations to the first equation in the BBGKY density-matrix hierarchy~\cite{LM18,LM20b}. Unlike most of the previously-discussed approaches, this approach respects initial-state dependence in the sense that the xc potentials for identical density evolutions arising from different initial states will differ. 

The simplest approximation would be to replace $\rho_1$ and $n\xc$ by their KS counterparts~\cite{RB09b,FNRM16,SLWM17,LSWM18,FLNM18,LHRC17}, an approximation that we dub $v\xc\S$. Although $v\xc\S$ generally  approximates the exact $v\xc^W$ well, the kinetic component vanishes $v\xc^T$, and it is the kinetic component that contains the large dynamical step and peak features~\cite{DLM22} (Sec.~\ref{sec:Hubbard}) that are crucial to accurately capture dynamics in a number of situations, e.g. electron scattering~\cite{SLWM17,LSWM18}, charge-transfer out of a ground state~\cite{FERM13} (see the example in Sec.~\ref{sec:Hubbard}), quasiparticle propagation through a wire~\cite{RG12}. 
Refs.~\cite{FLNM18} explored various ``frozen" approximations, while Ref.~\cite{LM18} presented an approach based on coupling the KS evolution to the first equation in the BBGKY hierarchy for the interacting 1RDM; the two equations ``pass" back and forth an approximation for a fictitious $\tilde\rho_2$ as a functional of the 1RDM evolving from the BBGKY equation and the KS 1RDM evolving from the KS equation. This was shown to capture the dynamical steps and peaks, to satisfy the ZFT and GTI,  be self-interaction-free, however it becomes numerically unstable after too short times to be practical~\cite{LM18,LM20b}. Whether instead a paradigmatic system can be found from which the interacting 1RDM can be obtained as a functional of KS-accessible quantities remains an open exploration.

\begin{figure}[htbp]
\includegraphics[width=0.45\textwidth]{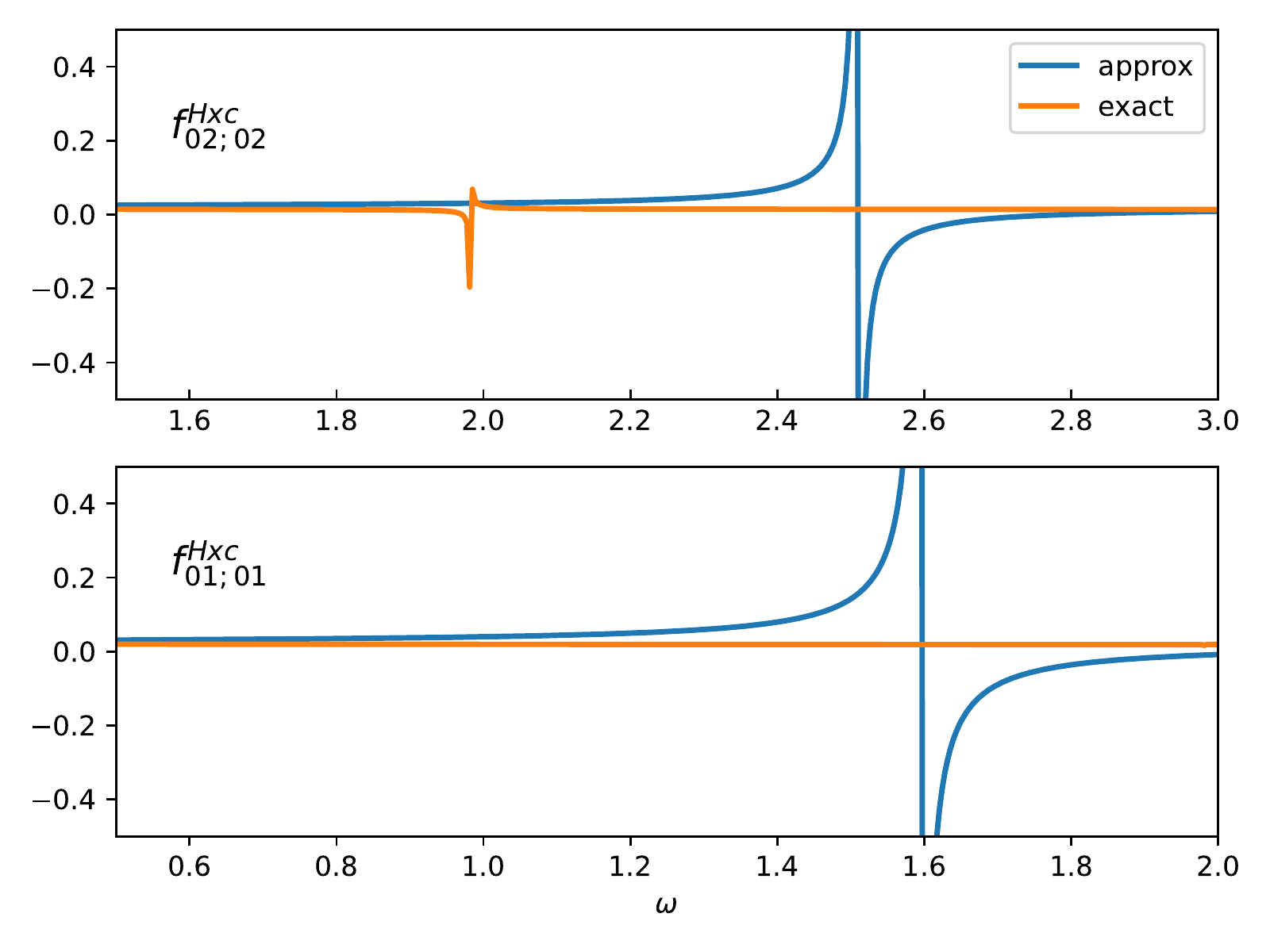} 
\caption{Frequency-dependence of the diagonal part of $f\Hxc$ in a model system (see \cite{MZCB04}) for the exact (orange) and density-matrix coupled approximation (blue). Upper panel shows the matrix element corresponding to a single excitation that couples to a double excitation, leading to a pole in the exact $f\Hxc$. Lower panel the matrix element corresponding to an uncoupled single excitation in the exact system.
}
\label{fig:fhxc}
\end{figure}

Even if the choices made so far for the 2RDM have led to a numerically unstable approach, a question is whether they capture non-adiabatic features  in the linear response domain related to phenomena such as double excitations. In particular, whether the kernel resulting from the functional derivative of $v\c^{\rm T}$ gives an approximation to the pole in the
exact kernel~\cite{MZCB04} that is responsible for the interaction of the singly and doubly-excited KS states underlying a state of double-excitation character (Sec.~\ref{sec:specific}).
This can be answered by deriving the linear response of the system starting with Eq.~(\ref{Eq:theo_vct}) and (\ref{Eq:theo_vxcw}) and transforming to the frequency-domain. In the KS basis, this creates a set of equations  to solve to obtain a kernel $f\Hxc(\omega)$ that can be used in Casida equations.
Testing this approach on the model studied in Ref.~\cite{MZCB04}, a one-dimensional harmonic oscillator with delta-interacting electrons, with the choice of $\tilde\rho_2$ as the time-dependent KS one, yields mixed results as seen in Fig.~\ref{fig:fhxc}. A pole does indeed appear in the kernel, showing a strong non-adiabaticity in the matrix element of $f\Hxc$ associated  with the state of double excitation character (upper panel). However, it has the wrong position, sign, and amplitude of the divergence. Moreover, a spurious pole also appears in the matrix element where the physical system only has a single excitation (lower panel). This shows the limitations of using  $\tilde\rho = \rho_{2{\sss S}}$ within the density-matrix coupled scheme.

\subsection{Specific cases}
\label{sec:specific}
While the previous sections propose universal xc functional approximations with memory, there have also been approximations derived for specific cases where memory is known to be important. We briefly mention some of these here.

As just discussed, the adiabatic approximation in linear response is unable to capture states with double excitation character~\cite{JCS96,TH00,MZCB04,CZMB04,EGCM11, C05,RSBS09, GB09,M22,PGR18}.  These states can enter the spectrum even at low energies, e.g. in conjugated polyenes where they mix with the single excitations~\cite{CZMB04}, and they are typically sprinkled throughout the spectrum in geometries away from equilibrium. This poses a problem for the general reliability of TDDFT predictions of photoinduced dynamics where nuclei are driven to sample a range of configurations due to their coupling to the electronic motion.  Inspired by the form of the exact xc kernel when a double excitation lies near a (group of) single excitation, Ref.~\cite{MZCB04} derived a frequency-dependent approximation to be applied for this case. This ``dressed" kernel has a pole in a frequency-range near the double excitation, and gave reasonable predictions for these states in a range of molecules~\cite{CZMB04,MW09,HIRC11}.  In a related spirit, a frequency-dependent quadratic response kernel was recently derived in Ref.~\cite{DRM22}, which cures unphysical divergences arising in adiabatic TDDFT when the difference between two excitation energies equals another one~\cite{PRF16}. 

A non-adiabatic approximation was derived for the single-impurity Anderson model~\cite{DSH18,DHK19}, capturing the dynamical step feature missing in the adiabatic approximation, and applied to quantum transport.  This functional depended only on the site occupation and its first time-derivative. Going from approximations on a lattice to real-space systems however is highly non-trivial:
as mentioned in Sec.~\ref{sec:Hubbard} even the basic theorems and $v$-representability issues are distinct on a lattice than in real-space, and whether the former can be  consistently converged to the latter is unclear.

Focussing on models where the ground-state is strongly-correlated, 
Refs.~\cite{TR17,ATZR20} derived an xc functional approximation for the xc kernel of Hubbard model systems from dynamical mean field theory; these could be applied to a real system with Hubbard parameters chosen somehow from experiment. The non-adiabatic part of the resulting kernel is completely local in space but has memory-dependence, placing it at risk of violating the zero force and Galilean invariance principles discussed earlier. 



\section{Outlook}

This review has focussed on non-adiabatic approximations to the xc potential or kernel, but another ingredient needed in TDDFT are functionals for the observables when the observables are not directly related to the density itself. For example, ionization probabilities~\cite{WB06}, momentum-distributions~\cite{WB07}, even simply the current-density whose rotational component is not generally reproduced by the TDKS system even if an exact xc functional is used~\cite{AV05,SK16,TK09b,GM12,DLFM21}. Usually these observables are extracted simply by taking expectation values of their usual operators in the KS state, which inherently entails an approximation additional to that of the xc functional. Corrections to such observable-functionals and their memory-dependent properties are largely unexplored.

The studies in the past years on the exact xc potential and its properties, and the different efforts in development of memory-dependent functionals summarized here show that the search for an accurate and practical non-adiabatic approximation is a challenging one.
The search is on-going and creative: several recent new directions not mentioned yet in this review have been proposed for time-dependent functional development which are still at a very preliminary stage, including coupling-constant integral transforms~\cite{G97,LM20b}, re-casting TDDFT using the second-time-derivative of the density as basic variable ~\cite{TU21}, and extensions of the connector theory approach~\cite{PGR18,VAPGR22} to the time-domain. Whether one of these will yield an elixir remains to be seen, but even if not, they reveal interesting physics about the dynamics of electron correlation.




\acknowledgements{
Financial support from the National Science Foundation Award CHE-2154829 (NTM)   and from the Department
of Energy, Office of Basic Energy Sciences, Division of Chemical
Sciences, Geosciences and Biosciences under Award No. DESC0020044, and the
 European Union's
Horizon 2020 research and innovation programme under the
Marie Sk{\l}odowska-Curie grant agreement No 101030447 (LL) are gratefully acknowledged.
 }
 
 \section{Competing interests}
 The Authors declare no Competing Financial or Non-Financial Interests. 

\section{Author Contributions}
N.T.M. conceived the project, summarized the literature, and wrote the majority of the manuscript. L.L. performed the calculations, analyzed the results, co-wrote  Sec.~\ref{sec:Hubbard} and Sec.~\ref{sec:DMxc} and contributed to the writing of all sections.  Both authors reviewed and edited the manuscript. 

\bibliography{./ref.bib}

\end{document}